# Optimistic Execution in Key-Value Store

Duong Nguyen
*Michigan State University*
nguye476@cse.msu.edu

Aleksey Charapko
*University at Buffalo, SUNY*
acharapk@buffalo.edu

Sandeep Kulkarni
*Michigan State University*
sandeep@cse.msu.edu

Murat Demirbas
*University at Buffalo, SUNY*
demirbas@buffalo.edu

***Abstract*—Limitations of CAP theorem imply that if availability is desired in the presence of network partitions, one must sacrifice sequential consistency, a consistency model that is more natural for system design. We focus on the problem of what a designer should do if she has an algorithm that works correctly with sequential consistency but is faced with an underlying key-value store that provides a weaker (e.g., eventual or causal) consistency. We propose a detect-rollback based approach: The designer identifies a correctness predicate, say $P$, and continue to run the protocol, as our system monitors $P$. If $P$ is violated (because the underlying key-value store provides a weaker consistency), the system rolls back and resumes the computation at a state where $P$ holds.**

**We evaluate this approach in the Voldemort key-value store. Our experiments with deployment of Voldemort on Amazon AWS shows that using eventual consistency with monitoring can provide $20-40\%$ increase in throughput when compared with sequential consistency. We also show that the overhead of the monitor itself is small (typically less than 8%) and the latency of detecting violations is very low. For example, more than $99.9\%$ violations are detected in less than $1$ second.**

## I. Introduction

Distributed key-value data stores have gained an increasing popularity due to their simple data model and high performance [1]. A distributed key-value data store, according to CAP theorem [2], [3], cannot simultaneously achieve sequential consistency and availability while tolerating network partitions. As network partition tolerance is considered as a must, it is inevitable to make trade-offs between availability and consistency, resulting in a spectrum of consistency models such as causal consistency and eventual consistency.

Weaker consistency models (e.g. causal, eventual [1], [4]–[9]) are attractive because they have the potential to provide higher throughput and higher customer satisfaction and they provide a platform on which users can build high performance distributed applications. On the other hand, weaker consistency models suffer from data conflicts. Although such data conflicts are infrequent [1], such incidences will effect the correctness of the computation and invalidate subsequent results.

On the other hand, developing algorithms for sequential consistency model is easier than developing those for weaker consistency models. Moreover, since sequential consistency model is *more natural*, the designer may already have access to an algorithm that is correct only under the sequential consistency. Thus, in this case, the question for the designer is what to do *if the underlying system provides weaker consistency* or *if the underlying system provides better performance under weaker consistency*?

Identify applicable funding agency here. If none, delete this.

As an illustration of such a scenario, consider a distributed computation that relies on a key-value store to arrange the exclusive access to a critical resource for the clients. If the key-value store employs sequential consistency, mutual exclusion is guaranteed [10], but the performance would be hurt due to the communication overhead of sequential consistency. If eventual consistency is adopted, then mutual exclusion is violated.

In this case, the designer has two options: (1) Either develop a brand new algorithm that works under eventual consistency, or (2) Run the algorithm by pretending that the underlying system satisfies sequential consistency but monitor it to detect violations of mutual exclusion requirement. In case of the first option, we potentially need to develop a new algorithm for every consistency model used in practice, whereas in case of the second option, the underlying consistency model is irrelevant although we may need to rollback the system to an earlier state if a violation is found. While the rollback in general distributed systems is a challenging task, existing approaches have provided rollback mechanisms for key-value stores with low overhead [11].

The predicate $P$ to monitor depends on the application. For the mutual exclusion application we alluded to above, $P$ might be concurrent access to the shared resource. As another example, consider the following. For many distributed graph processing applications, clients process a given set of graph nodes. Since the state of a node depends on its neighbors, clients need to coordinate to avoid processing two neighboring nodes simultaneously. In this case, predicate $P$ is the conjunction of smaller predicates proscribing the concurrent access to some pairs of neighboring nodes (Note that pairs of neighboring nodes belonging to the same client do not need monitoring). The system will continue executing as long as predicate $P$ is true. If $P$ is violated, it will be rolled back to an earlier state from where subsequent execution will continue (cf. Figure 1).

For performant execution, we require that the monitoring module is non-intrusive, i.e., it allows the underlying system to execute unimpeded. To evaluate the effectiveness of the module, we need to identify three parameters: (1) benefit of using the module instead of relying on sequential consistency,

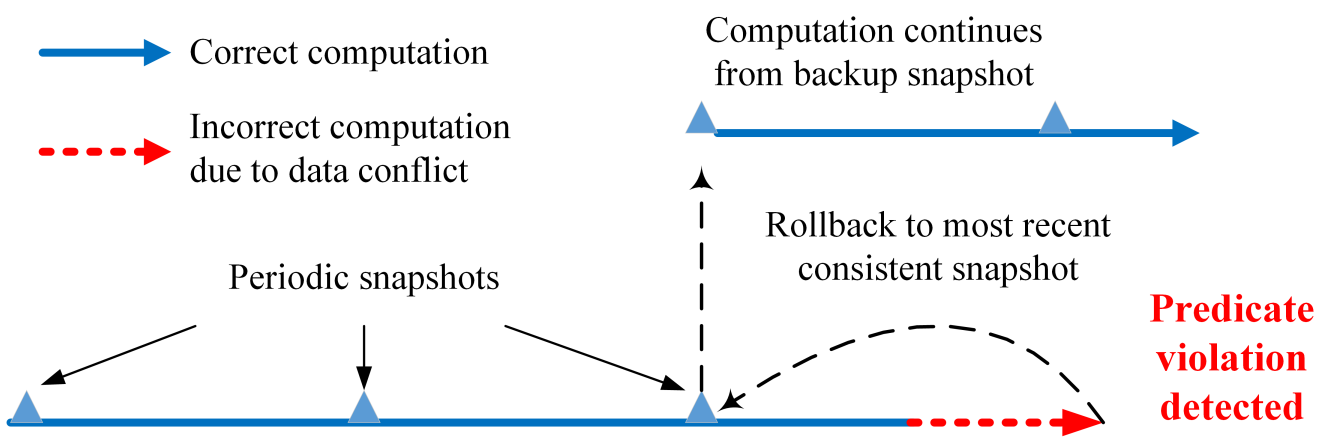

Fig. 1. An approach that support high performance and reliable distributed computations under weak consistency model. When data conflict occurs, the predicate of interest is violated. The violation is detected, system state is restored to the most recent snapshot and the computation would continue from there

(2) overhead of the module, i.e., how the performance is affected when we introduce the module, and (3) latency of the module, i.e., how long the module takes to detect violation of $P$. (Note that since the module is non-intrusive, it cannot prevent violation of $P$.)

**Contributions of the paper.** We implement a prototype for the predicate detection module for the Voldemort key-value data store and run experiments on the Amazon EC2 M2.large instances. For this key-value store, we develop monitoring algorithms for linear and semilinear predicates based on the algorithms in [12]–[14]. Our algorithms use Hybrid Vector Clock [15] to help saving resources from examining false positive cases thanks to its loosely synchronization with physical clock [16]. The observations from this work are as follows:

- We run simple graph distributed computations with mutual exclusion both on sequential consistency without the monitoring module and on eventual consistency with the monitoring module. We observe that –even with the overhead of the monitor– eventual consistency achieves a throughput 20% to 40% higher than that of sequential consistency. Furthermore, in those experiments, we find that violation of mutual exclusion is very rare. Hence, the cost of predicate detection and state rollback is outweighed by the benefit of a boosted throughput while the reliability of the computation is still preserved.
- We also evaluate the overhead of the monitoring module if it is intended solely for debugging or runtime monitoring. We find that when the monitor is used with sequential consistency, the overhead is at most 13%. And, for eventual consistency, the overhead is less than 11%.
- Regarding the latency of the module, 99.94% of violations are detected within 100 milliseconds, and in all cases are within thirteen seconds.

**Organization of the paper:** Section II we describe the architecture of the key-value store used in this paper. In section III, we define the notion of causality and identify how uncertainty of event ordering in distributed system affects the problem of predicate detection. Section IV describes the overall architecture of the system using monitors. Section V explains the structure of the predicate detection module used in this paper. Section VI presents experimental results and discussions. Section VII compares our paper with related work and we conclude in Section VIII.

## II. System Architecture

### A. Distributed Key-Value Store

We utilize the standard architecture for key-value stores. Specifically, the data consists of (one or more) tables with two fields, an unique key and the corresponding value. The field value consists of a list of $< version, value >$ pairs. A version is a vector clock that describes the origin of the associated value. It is possible that a key has multiple versions when different clients issue PUT (write) requests for that key independently. When a client issues a GET (read) request for a key, all existing versions of that key will be returned. The client could resolve multiple versions for the same key on its own or use the resolver function provided from the library. To provide efficient access to this table, it is divided into multiple partitions. Furthermore, to provide redundancy and ease of access, the table is replicated across multiple replicas.

To access the entries in this table, the client utilizes two operations, GET and PUT. The operation GET($x$) provides client the value associated with key $x$. And, the operation, PUT($x, val$), changes the value associated with key $x$ to $val$. The state of the servers can be changed only by PUT requests from clients.

### B. Voldemort Key Store

Voldemort is LinkedIn's open source equivalence of Amazon's Dynamo key-value store. In Voldemort, clients are responsible for handling replication. When connecting to a server for the first time, a client receives meta-data from the server. The meta-data contains the list of the servers and their addresses, replication factor ($N$), required reads ($R$), required writes ($W$), and other configuration information.

When a client wants to perform a PUT (or GET), it sends PUT (GET) requests to $N$ servers and waits for the responses for a predefined amount of time (timeout). If at least $W$ ($R$) acknowledgements (responses) are received before the timeout, the PUT (GET) request is considered successful. If not, the client performs one more round of request to other servers to get the necessary numbers of acknowledgements (responses). After the second round, if still less than $W$ ($R$) replies are received, the PUT (GET) request is not successful.

Since the clients do the task of replication, the values $N$, $R$, $W$ specified in the meta-data is only a suggestion. The clients can tune those values for their needs. By adjusting the value of $W$, $R$, and $N$, client can tune the consistency model. For example, if $W + R > N$ and $W > \frac{N}{2}$ for every client, then they will obtain sequential consistency. On the other hand, if $W + R \leq N$ then it is eventual consistency.

## III. The Problem of Predicate Detection

The goal of the predicate detection algorithm is to ensure that the predicate $P$ is always satisfied during the execution of the system. In other words, we want monitors to notify us of cases where predicate $P$ is violated.

Each process execution in a distributed system results in changing its local state, sending messages to other processes or receiving messages from other processes. In turn, this creates a partial order among local states of the processes in distributed systems. This partial order, happened-before relation [17], is defined as follows:

Given two local states $a$ and $b$, we say that $a$ happened before $b$ (denoted as $a \rightarrow b$) iff

- $a$ and $b$ are local states of the same process and $a$ occurred before $b$,
- There exists a message $m$ such that $a$ occurred before sending message $m$ and $b$ occurred after receiving message $m$, or
- There exists a state $c$ such that $a \rightarrow c$ and $c \rightarrow b$.

We say that states $a$ and $b$ are concurrent (denoted as $a||b$) iff $\neg(a \rightarrow b) \quad \wedge \quad \neg(b \rightarrow a)$

To detect whether the given predicate is violated, we utilize the notion of *possibility* modality [18], [19]. In particular, the goal is to find a set of local states $e_1, e_2, ..e_n$ such that

- One local state is chosen from every process,
- All chosen states are pairwise concurrent.
- The predicate $\neg P$ is true in the global state $\langle e_1, e_2, \cdots, e_n\rangle$

### A. Vector Clocks and Hybrid Vector Clocks

To determine whether state $a$ happened before state $b$, we can utilize vector clocks or hybrid vector clocks. Vector clocks, defined by Fidge and Mattern [20], [21], are designed for asynchronous distributed systems that make no assumption about underlying speed of processes or about message delivery. Hybrid vector clocks [15] are designed for systems where clocks of processes are synchronized within a given synchronization error (parameter $\epsilon$). While the size of vector clock is always $n$, the number of processes in the system, hybrid vector clocks have the potential to reduce the size to less than $n$.

Our predicate detection module can work with either of these clocks. For simplicity, we recall hybrid vector clocks (HVC) below.

Every process maintains its own HVC. HVC at process $i$, denoted as $HVC_i$, is a vector with $n$ elements such that $HVC_i[j]$ is the most recent information process $i$ knows about the physical clock of process $j$. $HVC_i[i] = PT_i$, the physical time at process $i$. Other elements $HVC_i[j], j \neq i$ is learned through communication. When process $i$ sends a message, it updates its HVC as follows: $HVC_i[i] = PT_i$, $HVC_i[j] = max(HVC_i[j], PT_i - \epsilon)$ for $j \neq i$. Then $HVC_i$ is piggy-backed with the outgoing message. Upon reception of a message $msg$, process $i$ will use the piggy-backed hybrid vector clock $HVC_{msg}$ to update its HVC: $HVC_i[i] = PT_i$, $HVC_i[j] = max(HVC_{msg}[j], PT_i - \epsilon)$ for $j \neq i$.

Hybrid vector clocks are vectors and can be compared as usual. Given two hybrid vector clock $HVC_i$ and $HVC_j$, we says $HVC_i$ is smaller than $HVC_j$, denoted as $HVC_i < HVC_j$, iff $HVC_i[k] \leq HVC_j[k] \forall k$ and $\exists l : HVC_i[l] < HVC_j[l]$. If $\neg(HVC_i < HVC_j) \wedge \neg(HVC_j < HVC_i)$, then the two hybrid vector clocks are concurrent, denoted as $HVC_i||HVC_j$.

If we set $\epsilon = \infty$, then hybrid vector clocks have the same properties as vector clocks. If $\epsilon$ is finite, certain entries in $HVC_i$ can have the default value $PT_i - \epsilon$ that can be removed. For example, if $n = 10, \epsilon = 20$, a hybrid vector clock $HVC_0 = [100, 80, 80, 95, 80, 80, 100, 80, 80, 80]$ could be represented by $n(10)$ bits 10010010001 and a list of three integers $100, 95, 100$, instead of a list of ten integers.

We use HVC in our implementation to facilitate its use when the number of processes is very large. However, in the experimental results we ignore this optimization and treat as if $\epsilon$ is $\infty$.

### B. Different Types of Predicate Involved in Predicate Detection

In the most general form, predicate $P$ is an arbitrary boolean function on the global state and the problem of detecting $\neg P$ is NP-complete [14]. However, for some types of predicates such as linear predicates, semilinear predicates, bounded sum predicates, there exist efficient detection algorithms [12]–[14]. In this paper, we adapt these algorithms for monitoring in key-value stores. Since the correctness of our algorithms follows from the existing algorithms, we omit detailed discussion of the algorithm and focus on its effectiveness in key-value stores.

## IV. A Framework for Optimistic Execution

The overall framework for optimistic execution in key-value store is as shown in Figure 2. In addition to the actual system execution in the key-value store, we include local detectors for every server. These local detectors provide information to the (one or more) monitors. Each monitor is designed to ensure that a property $P$ continues to be true during the execution. In other words, it is checking if a consistent snapshot where $\neg P$ is true.

When the monitor detects violation of the desired property $P$, it notifies the rollback module. The rollback module can stop or rollback the subsequent system execution and continue the system execution.

If violation of predicate $P$ is rare and the overall system execution is short, we could simply restart the computation from the beginning.

If the system computation is long, we can take periodic snapshots. Hence, when a violation is found, we can notify all clients and servers to stop the subsequent computation until the restoration to previous checkpoint is complete. The exact length of the period would depend upon the cost of taking the snapshot and probability of violating predicate $P$ in the interval between snapshots.

In case the violations are frequent, feedbacks from the monitor can help the clients to adjust accordingly. For example, if Voldemort clients are running in eventual consistency and find that their computations are restored too frequently, they can switch to sequential consistency by tuning the value of $R$ and $W$ without the involvement of the servers (Recall that in Voldemort key-value store, the clients are responsible for replication).

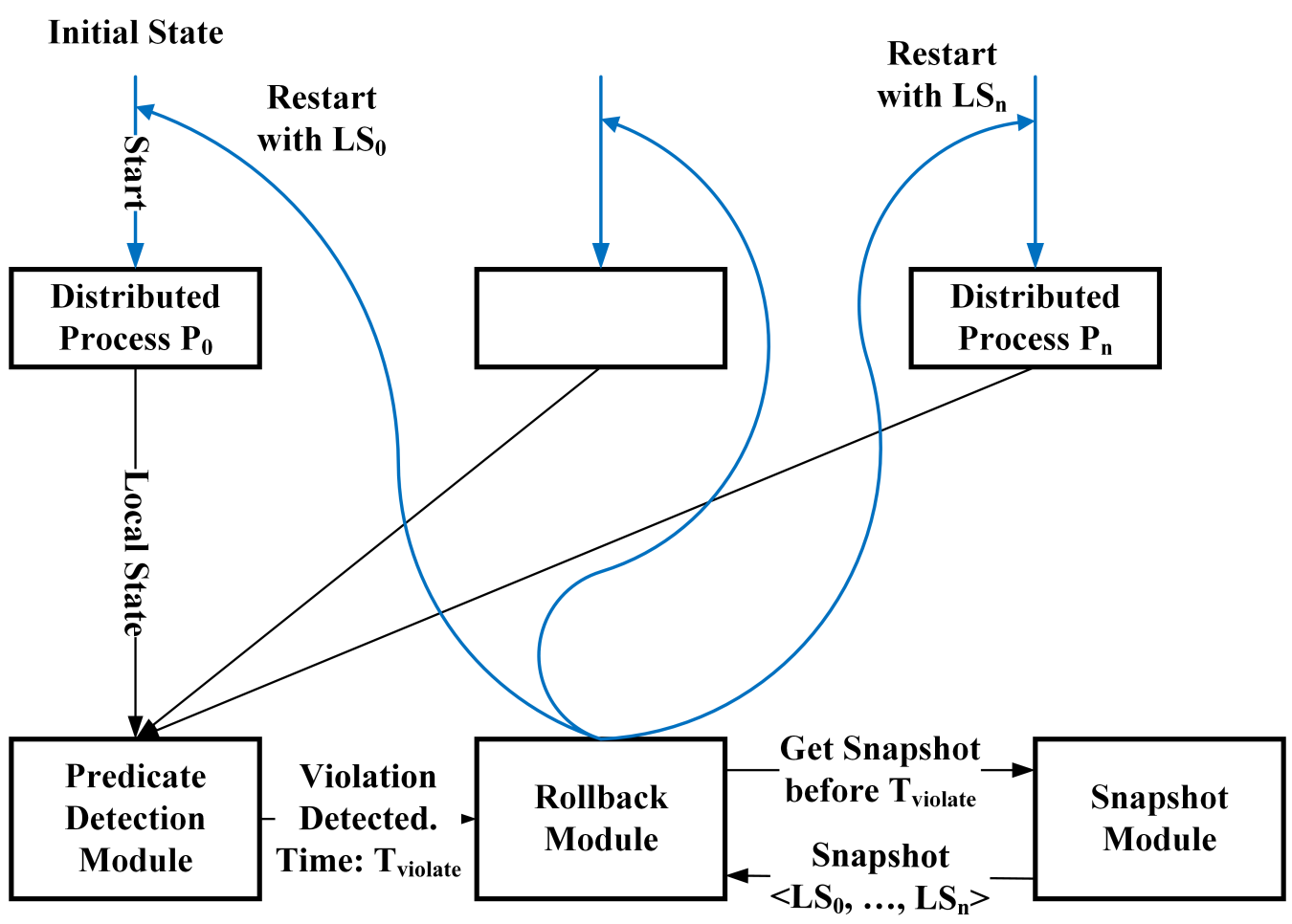

Fig. 2. An overall framework for optimistic execution in key-value store.

Alternatively, we can utilize approach such as Retroscope [11]. Once the violation is detected, the predicate detection module can identify a safe estimate of the start time $T_{violate}$ at which the violation occurred based on the timestamps of local states it received. Retroscope allows us to dynamically create a consistent snapshot that was valid just before $T_{violate}$ if $T_{violate}$ is within the window-log. This is possible if the predicate detection module is effective enough to detect the violation promptly. In [11], it authors have shown that it is possible to enable rollback for up to 10 minutes while keeping the size of logs manageable.

The approach in Retroscope can be further optimized by identifying the cause of the rollback. For example, consider the example from the Introduction that considers a graph application and requires that two clients do not operate on neighboring nodes simultaneously. Suppose a violation is detected due to clients $C_1$ and $C_2$ operating on neighboring nodes $V_1$ and $V_2$. In this case, we need to rollback $C_1$ and $C_2$ to states before they operated on $V_1$ and $V_2$. However, clients that do not depend upon the inconsistent values of nodes $V_1$ and $V_2$ need not be rolled back.

The goal of this paper is to evaluate the effectiveness of the monitor. In particular, our goal is to determine the overhead of such a monitor and the benefit one could get by running the algorithm with a weaker consistency model. Since this benefit is independent of the strategy used for rollback, we only focus on the effectiveness and overhead of the monitor. With this motivation, the properties of interest in this paper are

- How much overhead occurs when monitors are introduced? This will help us analyze the overhead when monitors are intended for debugging.
- How does the performance of the system compare under sequential consistency model (where $P$ was guaranteed to be true as the algorithm is correct) with the performance under a weaker consistency model with the monitor?
- How frequent are violations of $P$? This would identify the strategy that is suitable for roll back.
- How long does it take to detect violation of $P$? This would help determine whether logs would be sufficient to provide rollback using approaches such as those in [11].

## V. Predicate Detection Module

The predicate detection module is responsible for monitoring and detecting violation of global predicates in a distributed system. The structure of the module is as shown in Figure 4. It consists of local predicate detectors attached to each server and the monitors independent of the servers. The local predicate detector watches the state of its host server and sends information to the monitors. This is achieved by intercepting the PUT request(s) when they change variables that may affect the predicates being monitored. The monitors run predicate detection algorithm based on the information received to determine if the global predicates of interest is violated (cf. Figure 4).

Our predicate detection module can monitor multiple predicates simultaneously, each monitor for one predicate. However, they share the local detectors. In other words, there is one local detector for each server and the number of monitors is equal to the number of predicates being monitored. Note that in Figure 4, each monitor is depicted as one process. In implementation, a monitor may consists of multiple distributed processes collaborating to monitor a single predicate. For simplicity, each monitor is one process in this paper discussion.

The goal of the monitor is to ensure that the given predicate $P$ is always satisfied during the execution. We anticipate that the predicate of interest being monitored is a conjunctive predicate that captures all constraints that should be satisfied during the execution. In other words, $P$ is of the form $P_1 \wedge P_2 \wedge \cdots P_l$. The job of the monitor is to identify an instance where $P$ is violated, i.e., to determine if there is a consistent cut where $\neg P_1 \vee \neg P_2 \vee \cdots \neg P_l$ is true. For this reason, users provide the predicate being detected ($\neg P$) in a disjunctive normal form. We use XML format to represent the predicate. For example, the predicate $\neg P \equiv (x_1 = 1 \wedge y_1 = 1) \vee z_2 = 1$ in XML format is shown in Figure 3. Observe that this XML format also identifies the type of the predicate (conjunctive, semilinear, etc.) so that the monitor can decide the algorithm to be used for detection. In this paper, we implement the monitors based on the predicate detection algorithms in [13], [14].

**Implementation of Local Predicate Detectors.** Upon execution of a PUT request, the server calls the interface function `localPredicateDetector` which examines the state change and sends a message (also known as a candidate) to one or more monitors if appropriate. Note that not all state changes cause the `localPredicateDetector` to send candidates to the monitors. The most common example for this is when the changed variable is not relevant to the predicates being detected. Other examples depend upon the type of predicate being detected. As an illustration, if predicate

```xml
<predicate>
  <type>semilinear</type>
  <conjClause>
    <id>0</id>
    <var>
      <name>x2</name> <value>1</value>
    </var>
    <var>
      <name>y2</name> <value>1</value>
    </var>
  </conjClause>
  <conjClause>
    <id>1</id>
    <var>
      <name>z2</name> <value>1</value>
    </var>
  </conjClause>
</predicate>
```

Fig. 3. XML specification for $\neg P \equiv (x_1 = 1 \wedge y_1 = 1) \vee z_2 = 1$

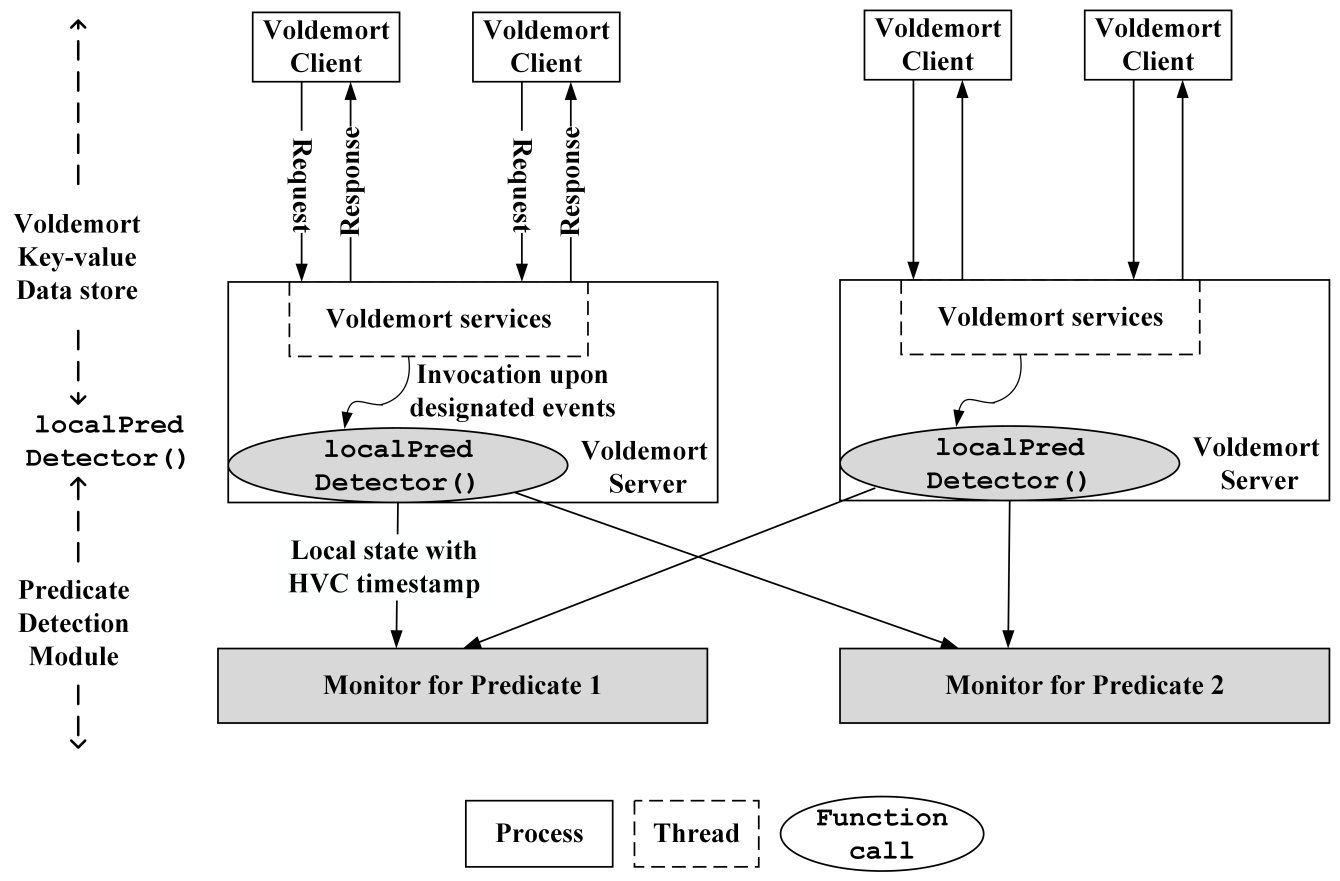

Fig. 4. Architecture of predicate detection module

$\neg P$ is of the form $x_1 \wedge x_2$ then we only need to worry about the case where $x_1$ changes from $false$ to $true$.

The local predicate detector maintains a cache of variables related to the predicates of interest to efficiently monitor the server state. A candidate sent to the monitor of predicate $P$ consists of an HVC interval and a partial copy of server local state containing variables relevant to $P$. The HVC interval is the time interval on the server when $P$ is violated, and the local state has the values of variables which make $\neg P$ become true.

For example, assume we want to detect when a global conjunctive predicate $\neg P_2 \equiv (\neg LP_2^1) \wedge (\neg LP_2^2) \wedge ...(\neg LP_2^n)$ becomes true. On server $i$, the local predicate detector will monitor the corresponding local predicate $\neg LP_2^i$ (or $\neg LP_2$ for short, in the context of server $i$ as shown in Figure 5). Since $\neg P_2$ is true only when all constituent local predicates are true, server $i$ only has to send candidates for the time interval when $\neg LP_2$ is true. In Figure 5, upon the first PUT request, no candidate is sent to monitor $M_2$ because $\neg LP_2$ is false during interval $[HVC_i^0, HVC_i^1]$. After serving the first PUT request, the new local state makes $\neg LP_2$ true, starting from the time $HVC_i^2$. Therefore upon the second PUT request, a candidate is sent to monitor $M_2$ because $\neg LP_2$ is true during the interval $[HVC_i^2, HVC_i^3]$. Note this candidate transmission is independent of whether $\neg LP_2$ is true or not after the second PUT request is served. It depends on whether $\neg LP_2$ is true after execution of the previous PUT request. That is why, upon the second PUT request, a candidate is also sent to monitor $M_3$ but none is sent to $M_1$. Note that if the predicate is semi-linear, then upon a PUT request for a relevant variable, the local predicate detector has to send a candidate to the associated monitor anyway.

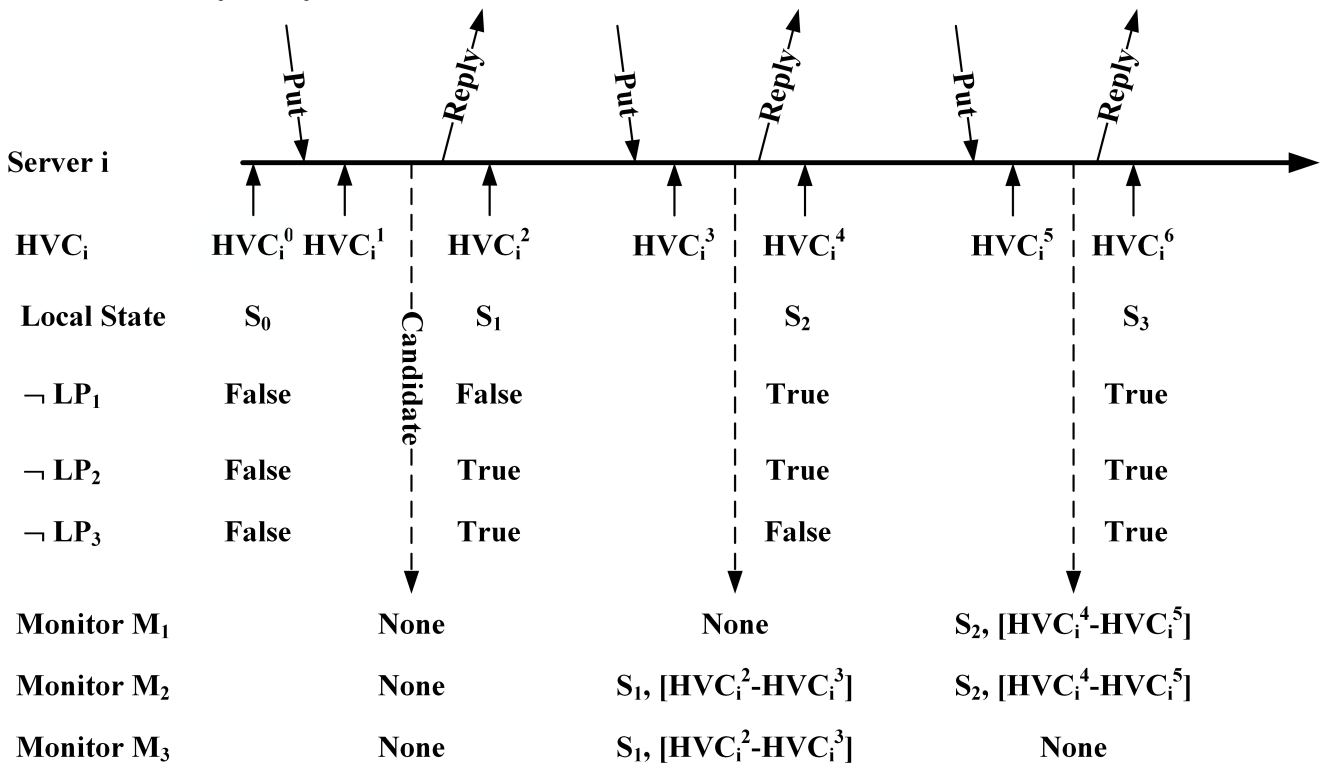

Fig. 5. Illustration of candidates sent from a server to monitors corresponding to three conjunctive predicates. If the predicate is semilinear, the candidate is always sent upon a PUT request of relevant variables.

**Implementation of the monitor.** The task of the monitor is to determine if the global predicate $P$ is violated, i.e., to detect if a consistent state with $\neg P$ exists in the system execution. The monitor constructs a global view of the variables relevant to $P$ from the candidates it receives. The global view is valid if there is no causal relationship [17] between the candidates.

The causal relationship between candidates is determined as follows: suppose we have two candidates $Cand_1, Cand_2$ from two servers $S_1, S_2$ and their corresponding HVC intervals $[HVC_1^{start}, HVC_1^{end}], [HVC_2^{start}, HVC_2^{end}]$. Without loss of generality, assume that $\neg(HVC_1^{start} > HVC_2^{start})$ (cf. Figure 6).

- If $HVC_2^{start} < HVC_1^{end}$ then the two intervals have common time segment and $Cand_1 \| Cand_2$.
- If $HVC_1^{end} < HVC_2^{start}$, and $HVC_1^{end}[S_1] \leq HVC_2^{start}[S_2] - \epsilon$ then interval one is considered happens before interval two. Note that $HVC[i]$ is the element corresponding to process $i$ in HVC. In this case $Cand_1 \rightarrow Cand_2$
- If $HVC_1^{end} < HVC_2^{start}$, and $HVC_1^{end}[S_1] > HVC_2^{start}[S_2] - \epsilon$, this is the uncertain case where the intervals may or may not have common segment. In order to avoid missing possible bugs, the candidates are considered concurrent.

Recall that for the global view to be valid, all HVC intervals must be pairwise concurrent.

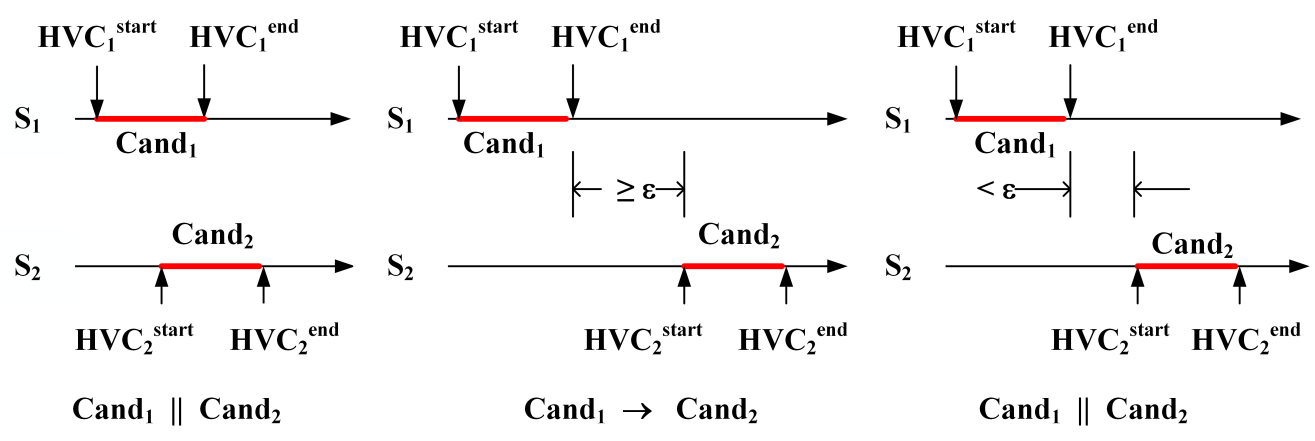

Fig. 6. Illustration of causality relation under HVC interval perspective

---

**Algorithm 1** Monitor algorithm for linear predicate

---

**Input:**
  $P$ ▷ global linear predicate to monitor
**Variable:**
  $GS$ ▷ global state
**Initialization:**
  $GS \leftarrow$ set of initial local states
**while** P(GS)==true **do**
  Find forbidden local state $s \in GS$
  $GS \leftarrow GS \cup succ(s)$ ▷ advance $GS$ along $s$
  consistent($GS$) ▷ make $GS$ consistent
**end while**
return $GS$

---

When a global predicate is detected, the monitor informs the administrator or triggers a designated process of recovery. We develop detection algorithms for the monitors of linear predicates and semilinear predicates based on [14] as shown in Algorithm 1 and Algorithm 2. Basically, the algorithms have to identify the correct candidates to update the global state ($GS$) so that we would not have to consider all possible combinations of $GS$ as well as not miss the possible violations. In linear (or semilinear) predicates, these candidates are forbidden (or semi-forbidden) states. Forbidden states are states such that if we do not replace them, we would not be able to find the violation. Therefore, we must advance the global state along forbidden states. Semi-forbidden states are states such that if we advance the global state along them, we would find a violation if there exists any. When advancing global state along a candidate, that candidate may not be concurrent with other candidates existing in the global state. In that case, we have to advance the candidates to make them consistent. This is done by `consistent(GS)` in the algorithm. If we can advance global state along a candidate without `consistent(GS)`, that candidate is called an eligible state. The set of all eligible states in global state is denoted as `eligible(GS)` in the algorithms. For more detailed discussion and proof of correctness of the algorithms, please refer to [12]–[14], [22].

## VI. Evaluation Results and Discussion

### A. Experimental Setup

We conducted our experiments on Amazon AWS M2.large instances that consist of two vCPUs and 8GB RAM. We considered replication factor ($N$) of 2, 3 and 5. The parameters

---

**Algorithm 2** Monitor algorithm for semilinear predicate

---

**Input:**
  $P$ ▷ global semilinear predicate to monitor
**Variable:**
  $GS$ ▷ global state
**Initialization:**
  $GS \leftarrow$ set of initial local states
**while** P(GS)==true **do**
  Find a local state $s \in GS$ such that $s \in eligible(GS)$ and $s$ a semi-forbidden state of $P$ in $GS$.
  $GS \leftarrow GS \cup succ(s)$ ▷ advance $GS$ along $s$
**end while**
return $GS$

---

$R$ (required reads) and $W$ (required writes) are chosen to achieve different consistency models as shown in Table I.

TABLE I
Setup of consistency models with $N$ (replication factor), $R$ (required reads), and $W$ (required writes)

| N | R | W | Abbreviation | Consistency model |
|---|---|---|---|---|
| 2 | 1 | 2 | N2R1W2 | Sequential |
| | 1 | 1 | N2R1W1 | Eventual |
| 3 | 1 | 3 | N3R1W3 | Sequential |
| | 1 | 1 | N3R1W1 | Eventual |
| 5 | 1 | 5 | N5R1W5 | Sequential |
| | 1 | 1 | N5R1W1 | Eventual |

The number of servers is equals to $N$. The number of clients is double the number of servers. Note that each server only has two virtual CPU, and the clients in our experiment keep requesting the server, thus the servers are employed at full capacity. Only in some experiments designed to stress the monitors, the number of clients can be four times as many as the number of servers. Monitors are distributed among the machines running the servers. We have done so to ensure that the cost of the monitors is accounted for in experimental results while avoiding overloading a single machine. An alternative approach is to have monitors on a different server. In this case, the trade-off is between CPU cycles used by the monitors (when monitors are co-located with servers) and communication cost (when monitors are on a different machine). Our experiments suggest that the latter (monitors on a different machine) is more efficient. However, since there is no effective way to compute the increased cost (of machines in terms of money), we have chosen to run monitors on the same machines as the servers.

We consider two types of predicates for monitoring. Our first application is motivated by graph applications. Here, we envision a graph G = (V, E) that consists of $V$ vertices and $E$ edges. The computation consists of a loop where in each loop, each client updates the state of a subset of vertices. To update the state of the vertex the client utilizes its own state and the state of its neighbors. To ensure that a vertex is updated correctly, it is necessary that while a client $C_1$ is working on node $V_1$, no other client is working on the neighbors of $V_1$. The goal of the monitors in this experiment is to detect violation of this requirement. When this requirement is violated, the monitors need to report and restore the system to a state before simultaneous updates to neighboring nodes were made. The number of predicates being monitored in this application is proportional to the number of edges of the form $(v_1, v_2)$ where different clients operate on $v_1$ and $v_2$.

The second predicate is a conjunctive predicate. This is a synthetic workload where the predicate being detected (i.e., $\neg P$) is of the form $P_1 \wedge P_2 \wedge \cdots \wedge P_l$. Each local predicate becomes true with a probability $\beta$ and the goal of the monitors is to determine if the global conjunctive predicate becomes true. Since we can control how frequently this predicate becomes true, we can use it mainly to detect monitoring latency and stress test the monitors.

### *B. Comparison of Sequential Consistency and Monitors with Eventual Consistency*

As discussed in the introduction, one of the problems faced by the designers is that they have access to an algorithm that is correct under sequential consistency but the underlying key-value store provides a weaker consistency. In this case, one of the choices is to pretend as if sequential consistency is available but monitor the critical predicate $P$. If this predicate is violated, we need to rollback to an earlier state and resume computation. Clearly, this approach would be feasible if the monitored computation with eventual consistency provides sufficient benefit compared with sequential consistency. In this section, we evaluate this benefit.

Figures 7 compares the performance of our algorithms for eventual consistency with monitor and sequential consistency without monitor with $N = 2, 3, 5$ and the percentage of PUT requests are $10\%, 25\%, 50\%$. From these figures, we find that when the percentage of PUT requests is small, the benefit is small. However, as the percentage of PUT requests increases, the benefit of eventual consistency with monitor increases. That is because in sequential setting, we emphasize $W = N$ over $R = 1$, when the percentage of PUT requests increases, the throughput of sequential consistency will decrease. On the other hand, in eventual consistency, the throughput is only slightly decreased because the the average work in PUT requests is more than the average work in GET requests. However, since eventual consistency utilizes $W = 1$, the effect of increased PUT requests is less.

When PUTs constitute half of the requests, the benefit of our algorithms when $N = 2, 3, 5$ is $25\%, 23\%, 36\%$, respectively. In other words, in these scenarios, our monitoring algorithm provides approximately 20-40% increase in throughput if we utilize eventual consistency and monitor it for possible violation.

### *C. Evaluating the Overhead of Monitoring*

Section VI-B considered the case where monitors are used to allow the application to use a key-value store with lower level of consistency. In this case, the goal of the monitors was to provide sufficient improvement in performance so that it is worthwhile to pay the penalty of occasional rollback.

Another application of monitoring is debugging to ensure that the program satisfies the desired property throughout the execution. For this case, we evaluate the effect of the overhead of monitoring by comparing the throughput with and without the monitors.

Figure 8 shows the overhead of the monitoring for eventual consistency in the graph application. We observe that the overhead of the monitor for $N = 2, 3$ and 5 is $2\%, 0.9\%$ and $4\%$ respectively. Note that in graph application, the number of predicates being monitored is proportional to the number of clients. Thus, the overhead remains reasonable even with monitoring several predicates simultaneously.

To further test the overhead of our algorithms, we ran experiments to detect a conjunctive predicate with 2 and 4 clients per server (cf. Figure 9, Table II). In this case, the monitor has to detect violation of a conjunctive predicate $P = P_1 \wedge P_2 \wedge \cdots P_{10}$. Furthermore, we can control how often these predicates become true by randomly changing when local predicates are true.

In these experiments, the rate of local predicate being true ($\beta$) is $1\%$, which is chosen based on the time breakdown of some MapReduce applications [23], [24]. We considered both eventual consistency and sequential consistency.

The overhead of our algorithms is smaller on eventual consistency than on sequential consistency because the number of requests generated to the monitor in case of eventual consistency is generally lower than that on sequential consistency. As the number of clients increases, the number of requests increases and thus the overhead of our algorithms increases. However, in our various experiments, the overhead is typically less than $8\%$. In experiments where each server is heavily loaded with 4 clients, the overhead does not exceed $13\%$.

Observing Figures 7, 8, 9, we find that there are a few moments where the aggregated throughput of all servers drops down. This is happening because some or all servers are spending a significant computation for the local predicate detection module. Such moments are infrequent. Furthermore, since each M2.large server used in our experiment has only two Voldemort server processes, when one of the process is running the predicate detection module, the aggregated throughput would be clearly affected. In a typical setting, each server runs a large number of server processes. Thus, when a server process is running predicate detection module, the decrease in aggregated throughput would be less noticeable.

### *D. Detection Latency*

In this section, we discuss the detection latency, i.e., the time elapsed between violation of the predicate being monitored and the time when the monitor detects it. Subsequently, we analyze its impact on potential approaches to rollback when predicate violation is detected.

In terms of violations of mutual exclusion requirement, we find that the violations are very rare. For example, during our various runs with $N = 2, 3, 5$, we only detected one instance

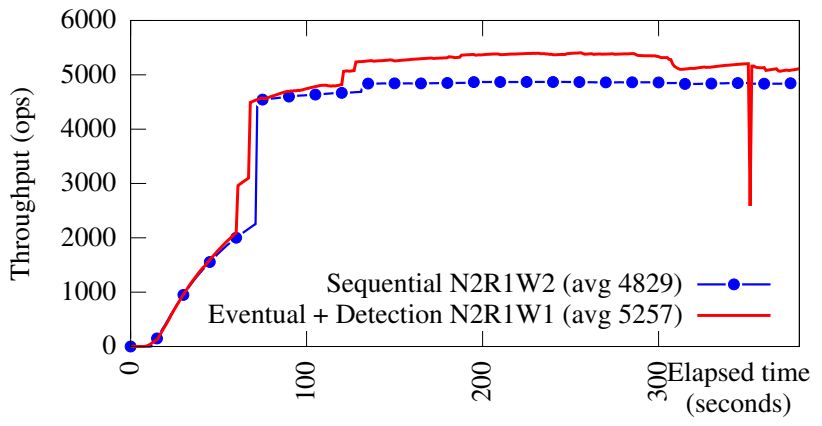

(a) N=2, PUT=10%, benefit=9%

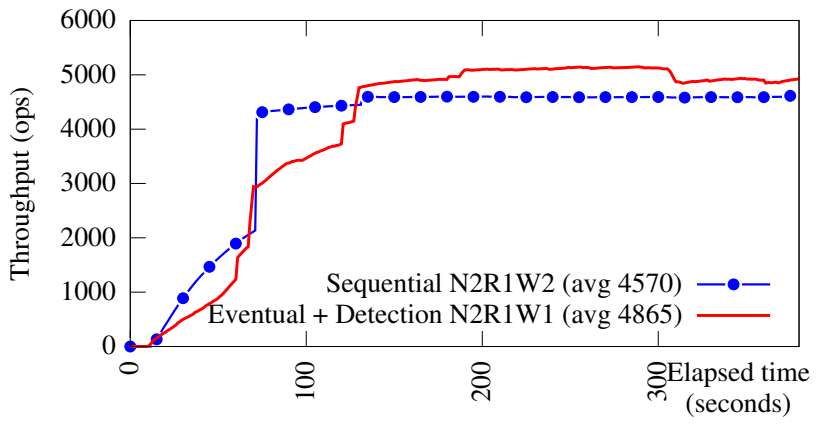

(b) N=2, PUT=25%, benefit=6%

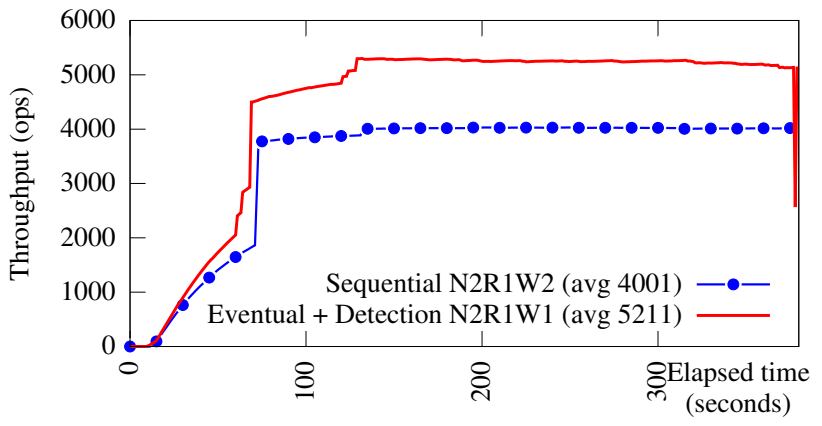

(c) N=2, PUT=50%, benefit=25%

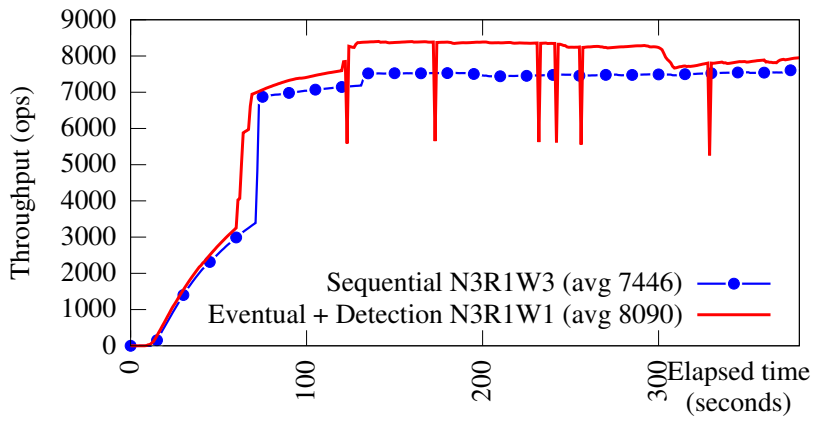

(d) N=3, PUT=10%, benefit=9%

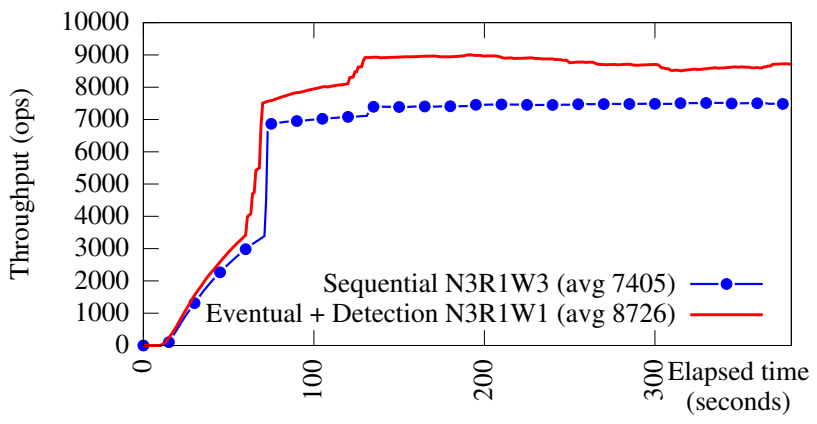

(e) N=3, PUT=25%, benefit=18%

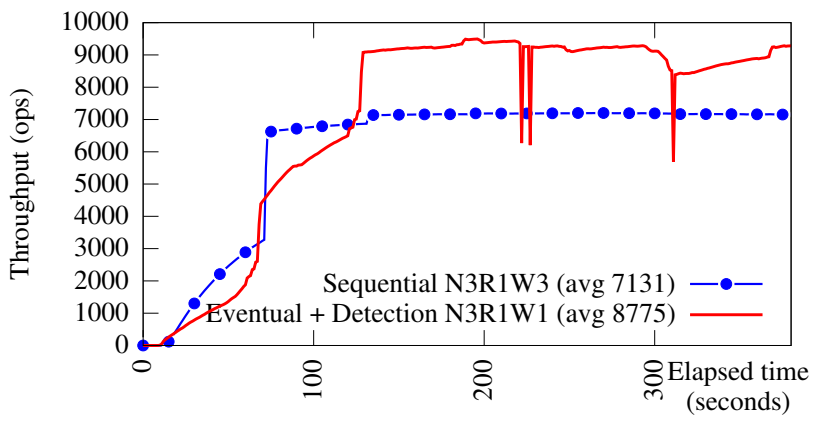

(f) N=3, PUT=50%, benefit=23%

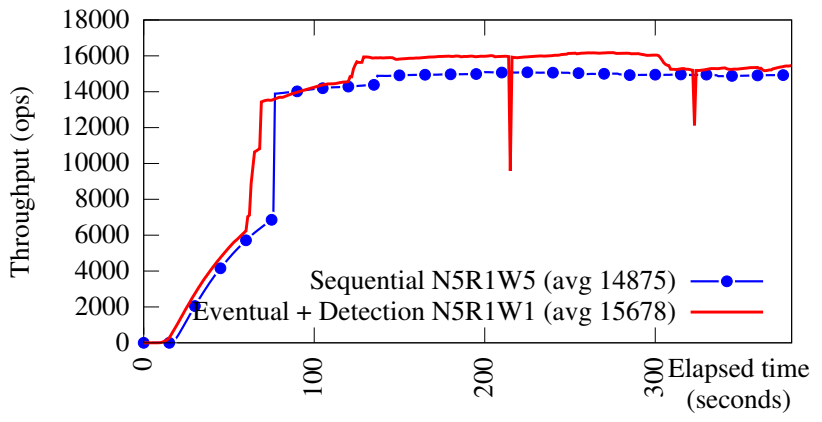

(g) N=5, PUT=10%, benefit=5%

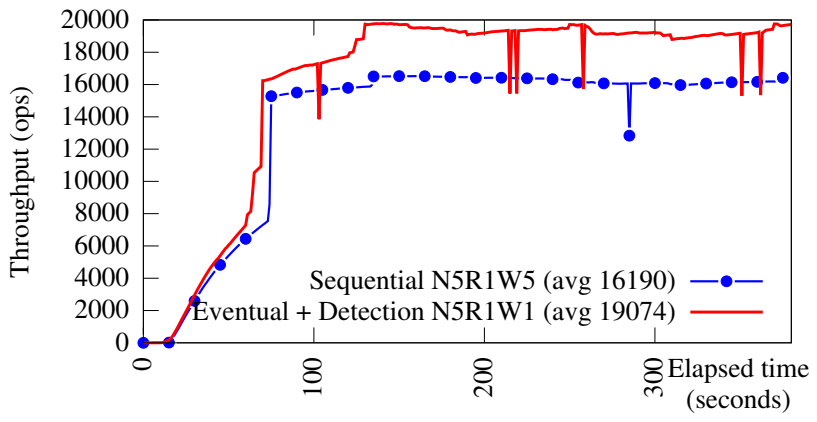

(h) N=5, PUT=25%, benefit=18%

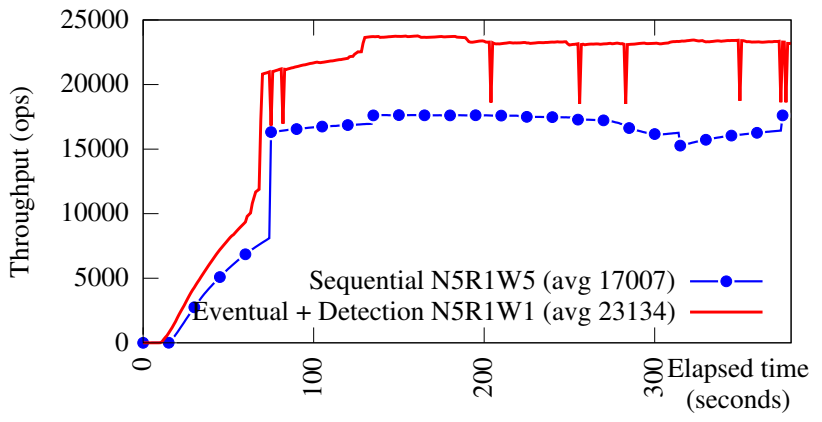

(i) N=5, PUT=50%, benefit=36%

Fig. 7. The benefit of Eventual Consistency + Predicate Detection vs. Sequential consistency in Graph Application. Number of clients per server is 2.

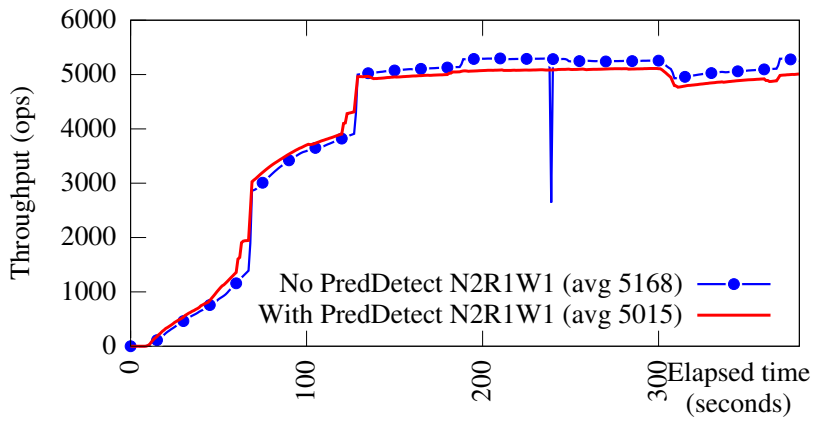

(a) N=2. Overhead=2%

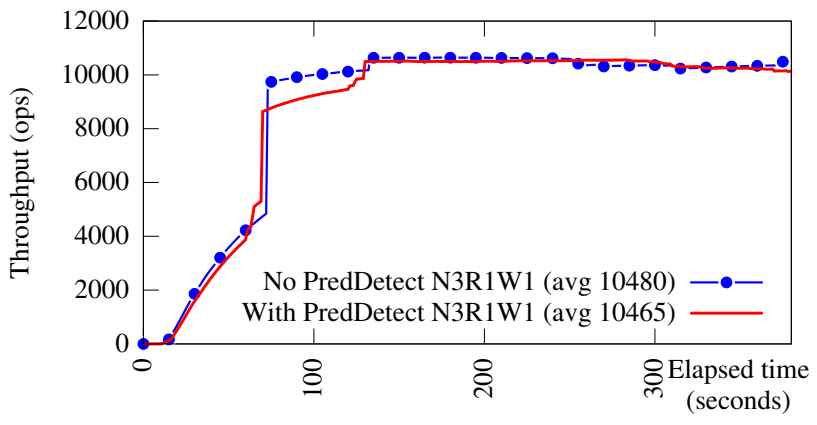

(b) N=3. Overhead=0.9%

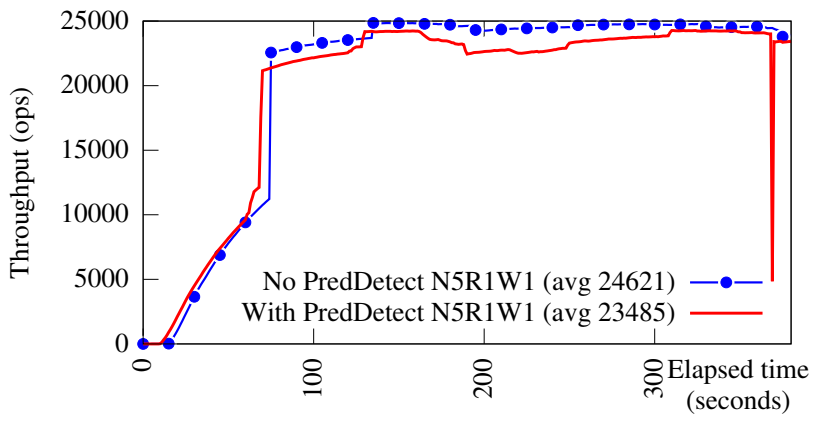

(c) N=4. Overhead=4%

Fig. 8. . The overhead of Predicate Detection on Eventual Consistency in *graph* application. Number of servers ($N$) = $2, 3, 5$. Number of clients per server is 2. Percentage of PUT requests is 50%

of mutual exclusion violation. This is because although violation of mutual exclusion in eventual consistency is a theoretical possibility, it is still rare. In that run (with $N = 3$), each client worked on (on average) 2098 nodes before a violation was found. Furthermore, when the violation occurred, it was detected within 20 milliseconds. Hence, we can easily utilize approaches such as Retroscope [11] to restore the system to a state before clients started working on conflicting nodes.

If we evaluate the overall violation frequency, it is even rare. In among all our experiments, clients operated on more than $40,000$ nodes. Experiments in the similar Dynamo key-value store with different applications also report the data inconsistency rate as low as $0.06\%$ [1]. This suggests that violations in eventual consistency are rare. And, in many applications, simply beginning from scratch may be a viable option. Furthermore, the detection latency is the property of the monitor. Specifically, this latency is independent of the actual workload provided to the monitor. Since this latency is very small, it would be possible to use Retroscope with just a few MBs of storage for rollback.

Regarding conjunctive predicates, we designed the experiments in such a way that the number of violations is large. Table III shows detection time distribution of more than 1.7 million violations recorded in the conjunctive predicate experiments. Predicate violations are generally detected within a second. Specifically, $99.94\%$ of violations were detected in $100ms$, $99.96\%$ of violations were detected in $1s$, $99.97\%$ of violations are detected in $2s$. There are a few cases where detection time is greater than ten seconds. Among all the runs, the maximum detection time recorded is 13 seconds.

## VII. Related Work

### A. Predicate Detection in Distributed Systems

Predicate detection is an important task in distributed debugging. An algorithm for capturing consistent global snapshots and detecting stable predicates was proposed by Chandy and

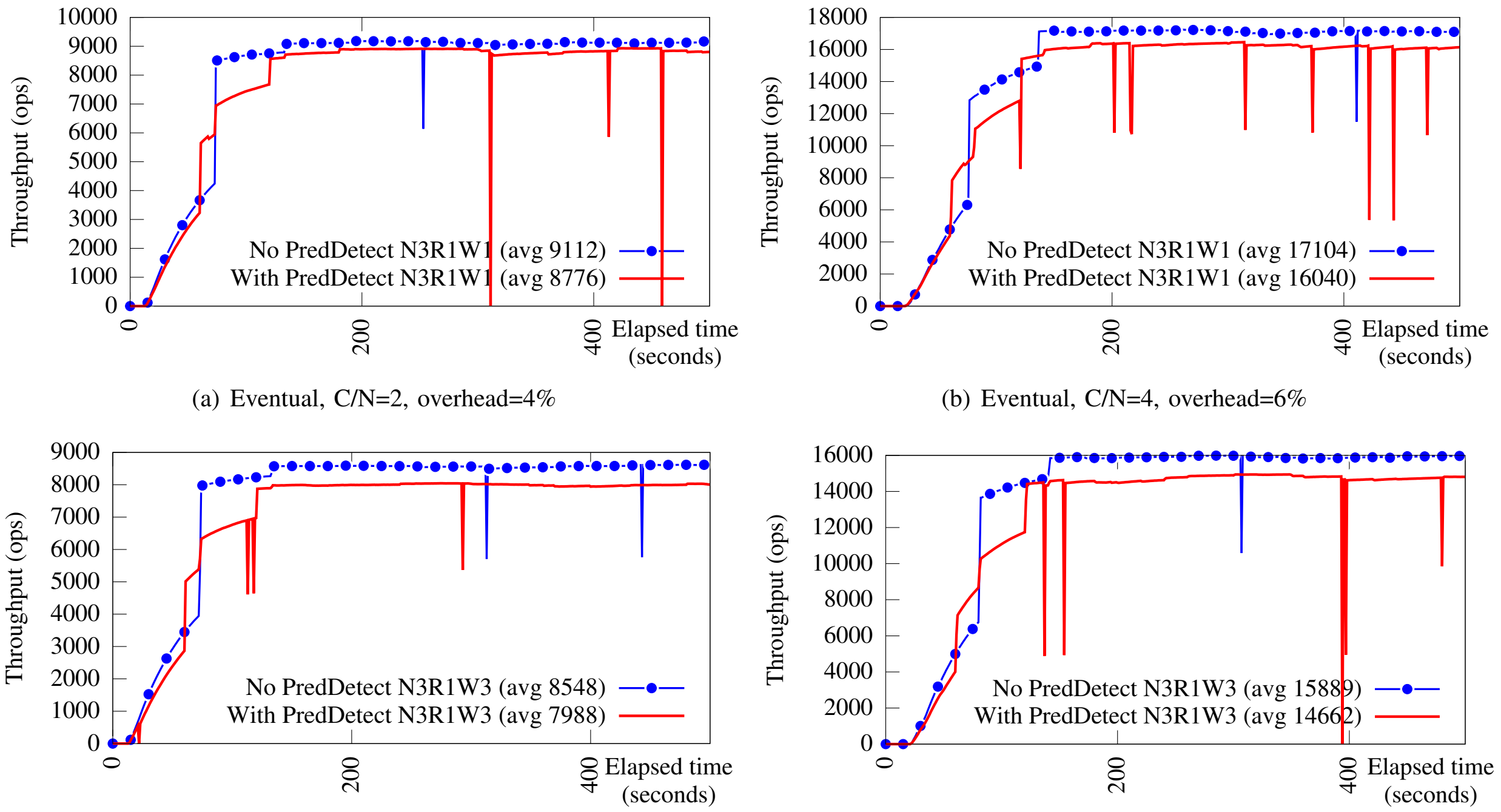

(a) Eventual, C/N=2, overhead=4%

(b) Eventual, C/N=4, overhead=6%

(c) Sequential, C/N=2, overhead=7%

(d) Sequential, C/N=4, overhead=8%

Fig. 9. The overhead of predicate detection on Eventual consistency (N=3, R=1, W=1) and Sequential consistency (N=3, R=1, W=3) in the *conjunctive* application. Percentage of PUT is 10%. Number of servers = 3, Number of clients per server ($C/S$) is 2 and 4. The overhead on Eventual consistency is smaller than on Sequential consistency. The overhead increases as the number of clients increases.

TABLE II
OVERHEAD OF PREDICATE DETECTION MODULE IN CONJUNCTIVE APPLICATION

| | | | N5R1W1 | N5R1W1 | N5R1W5 | N5R1W5 | N3R1W1 | N3R1W1 | N3R1W3 | N3R1W3 |
|---|---|---|---|---|---|---|---|---|---|---|
| Clients/server | Put | Monitor | Thru/put | Overhead | Thru/put | Overhead | Thru/put | Overhead | Thru/put | Overhead |
| 2 | 0.1 | Yes | 17114 | 8% | 15188 | 7% | 9112 | 4% | 8548 | 7% |
| | | No | 15911 | | 14204 | | 8776 | | 7988 | |
| | 0.25 | Yes | 20981 | 6% | 16779 | 7% | 9940 | 5% | 8157 | 8% |
| | | No | 19720 | | 15682 | | 9480 | | 7574 | |
| | 0.5 | Yes | 25352 | 8% | 17694 | 4% | 10744 | 5% | 7759 | 4% |
| | | No | 23462 | | 17046 | | 10219 | | 7451 | |
| | | | | | | | | | | |
| 4 | 0.1 | Yes | 31098 | 11% | 28143 | 13% | 17104 | 7% | 15889 | 8% |
| | | No | 27918 | | 25003 | | 16040 | | 14662 | |
| | 0.25 | Yes | 37644 | 10% | 30716 | 11% | 18423 | 8% | 15309 | 6% |
| | | No | 34193 | | 27588 | | 17009 | | 14378 | |
| | 0.5 | Yes | 43062 | 7% | 32751 | 11% | 19854 | 10% | 14805 | 8% |
| | | No | 40137 | | 29544 | | 18064 | | 13770 | |

TABLE III
RESPONSE TIME IN 1701805 CONJUNCTIVE PREDICATE VIOLATIONS

| Response time (milliseconds) | Count | Percentage |
|---|---|---|
| $< 100$ | 1700847 | 99.944% |
| $100 - 1000$ | 308 | 0.018% |
| $1000 - 2000$ | 244 | 0.014% |
| $2000 - 14000$ | 406 | 0.024% |

Lamport [25]. A framework for general predicate detection is introduced by Marzullo and Neiger [19] for asynchronous systems, and Stollers [18] for partially synchronous systems. These general frameworks face the challenge of state explosion as the predicate detection problem is NP-hard in general [14]. However, there exist efficient detection algorithms for several classes of practical predicates such as unstable predicates [22], [26], [27], conjunctive predicates [13], [28], linear predicates, semilinear predicates, bounded sum predicates [14]. Some techniques such as partial-order method [29] and computation slicing [30], [31] are also approaches to address the NP-Completeness of predicate detection. Those works use vector clocks to determine causality and the monitors receive states directly from the constituent processes. Furthermore, the processes are static. [32], [33] address the predicate detection in dynamic distributed systems. However, the classes of predicate is limited to conjunctive predicate. In this paper, our algorithms are adapted for detecting the predicate from only the states of the servers in key-value store, not from the clients.

The servers are static (except failure), but the clients can be dynamics. The predicates supported include linear (including conjunctive) predicates and semilinear predicates.

We use hybrid vector clocks to determine causality in our algorithms. In [16], the authors discussed the impact of various factors, among which is clock synchronization error, on precision of predicate detection module. In this paper, we set epsilon at a safe upper bound for practical clock synchronization error to avoid missing potential violations. In other words, hybrid vector clock is practically vector clock. Furthermore, this paper focuses on the efficiency and effectiveness of predicate detection module.

### B. Distributed data-stores

Many NoSQL data-stores exist on the market today, and a vast portion of these systems provide eventual consistency. The eventual consistency model is especially popular among key-value and column-family databases. The original Dynamo [1] was one of pioneers in the eventual consistency movement and served as the basis for Voldemort key-value store. Dynamo introduced the idea of hash-ring for data-sharding and distribution, but unlike Voldemort it relied on server-side replication instead of active client replication. Certain modern databases, such as Cosmos DB and DynamoDB [34], [35] offer tunable consistency guarantees, allowing operators to balance consistency and performance. This flexibility would enable some applications to take advantage of optimistic execution, while allowing other applications to operate under stronger guarantees if needed. However, many data-stores [36], [37] are designed to provide strong consistency and may not benefit from optimistic execution module.

Aside from general purpose databases, a variety of specialized solutions exist. For instance, TAO [38] handles social graph data at Facebook. TAO is not strongly consistent, as its main goal is performance and high scalability, even across datacenters and geographical regions. Gorilla [39] is another Facebook's specialized store. It operates on performance time-series data and highly tuned for Facebook's global architecture. Gorilla also favors availability over consistency in regards to the CAP theorem.

### C. Snapshots and Reset

The problem of acquiring past snapshots of a system state and rolling back to these snapshots has been studied extensively. Freeze-frame file system [40] uses Hybrid Logical Clock (HLC) to implement a multi-version Apache HDFS. Retroscope [11] takes advantage of HLC to find consistent cuts in the system's global state be examining the state-history logs independently on each node of the system. The snapshots produced by Retroscope can later be used for node reset by simple swapping of data-files. Eidetic systems [41] take a different approach and do not record all prior state changes. Instead, eidetic system records any non-deterministic changes at the operating system level and constructing a model to navigate deterministic state mutations. This allows the system to revert the state of an entire machine, including the operating system, data and applications, to some prior point. Certain applications may not require past snapshots and instead need to quickly identify consistent snapshots in the presence of concurrent requests affecting the data. VLS [42] is one such example designed to provide snapshots for data-analytics applications while supporting high throughput of requests executing against the system.

## VIII. Conclusion

Due to limitations of CAP theorem and the desire to provide availability/good performance during network partitions (or long network delays), many key-value stores choose to provide a weaker consistency such as eventual or causal consistency. This means that the designers need to develop new algorithms that work correctly under such weaker consistency model. An alternative approach is to run the algorithm by ignoring that the underlying system is not sequentially consistent but monitor it for violations that may affect the application. For example, in case of graph-based applications (such as those encountered in weather monitoring, social media analysis, etc.), each client operates on a subset of nodes in the graph. It is required that two clients do not operate on neighboring nodes simultaneously. In this case, the predicate of interest is that local mutual exclusion is always satisfied.

We demonstrated the usage of this approach in Voldemort in cases where we have two types of predicates: conjunctive predicates and semilinear predicates (such as that required for local mutual exclusion). We find that under a typical execution in Amazon AWS, we get a benefit of about $20-40\%$ increased throughput when we run eventual consistency with monitor. Furthermore, the violation of the predicate of interest in this graph-based application was very rare.

Furthermore, it is also feasible to utilize these monitors for the sake of debugging as well. In particular, the overhead of the monitor by itself is very low. The overhead is typically less than $8\%$ and in stressed experiments less than $13\%$.

The monitors detect violation quickly. In our experiments, we found that $99.94\%$ of violations were detected under $100ms$, $99.96\%$ violations were detected in $1s$. Only $0.024\%$ violations took more than $2s$. Thus, the amount of work wasted due to rollback would be very small especially if one utilizes techniques such as Retroscope [11] that allows one to roll back the system to an earlier state on-demand.

There are several possible future work in this area. This paper considered conjunctive and semilinear predicates. In general, the problem of predicate detection is NP-complete. Hence, we intend to evaluate the practical cost of these algorithms. We are also working on making these algorithms more efficient by permitting them to occasionally detect phantom violations. We are evaluating whether this increased efficiency would be worthwhile even though some unnecessary rollbacks may occur. Another future work is to integrate the monitor with Retroscope [11] to automate the rollback and recovery.